\documentclass[preprint,aps,superscriptaddress,nofootinbib,tightenlines,floats,floatfix]{revtex4}

\usepackage{epsf}

\newcommand{\beq}{\begin{equation}}
\newcommand{\eeq}{\end{equation}}
\newcommand{\beqa}{\begin{eqnarray}}
\newcommand{\eeqa}{\end{eqnarray}}

\def\OMIT#1{{}}
\def\Bbar{\overline{B}{}}

\newcommand{\ASL}{A_{\rm SL}}
\newcommand{\aSL}{a_{\rm SL}}
\renewcommand{\Re}{\mbox{Re}}
\renewcommand{\Im}{\mbox{Im}}
\newcommand{\no}{\nonumber}
\def\lsim{\lesssim}

\def\rhobar{\bar\rho}
\def\etabar{\bar\eta}
\newcommand{\ov}{\overline}

\begin{document}

\preprint{\vbox{ \hbox{LAL 02--01} 
  \hbox{LBNL--49436} \hbox{WIS/4/02--Jan--DPP} \hbox{hep-ph/0202010} }}

\title{\boldmath Implications of the CP asymmetry in semileptonic $B$ decay}

\vspace*{1.5cm}

\author{Sandrine Laplace\vspace{10pt}}
\affiliation{Laboratoire de l'Acc\'el\'erateur Lin\'eaire
	IN2P3-CNRS et Universit\'e Paris-Sud\\
	BP 34, F-91898 Orsay Cedex, France\vspace{6pt}}

\author{Zoltan Ligeti}
\affiliation{Ernest Orlando Lawrence Berkeley National Laboratory\\
        University of California, Berkeley, CA 94720\vspace{6pt}}

\author{Yosef Nir}
\affiliation{Department of Particle Physics, Weizmann Institute of Science\\
	Rehovot 76100, Israel\\ $\phantom{}$}

\author{Gilad Perez}
\affiliation{Department of Particle Physics, Weizmann Institute of Science\\
	Rehovot 76100, Israel\\ $\phantom{}$}

\begin{abstract} \vspace*{.2cm}

Recent experimental searches for $\ASL$, the CP asymmetry in semileptonic $B$
decay, have reached an accuracy of order one percent. Consequently, they give
meaningful constraints on new physics. We find that cancellations between the
Standard Model (SM) and new physics contributions to  $B^0-\Bbar^0$ mixing
cannot be as strong as was allowed prior to these measurements. The predictions
for this asymmetry within the SM and within models of minimal flavor violation
(MFV) are below the reach of present and near future measurements. Including
order $m_c^2/m_b^2$ and $\Lambda_{\rm QCD}/m_b$ corrections we obtain the SM
prediction: $-1.3\times10^{-3} < A_{\rm SL} < -0.5\times10^{-3}$.  Future
measurements can exclude not only the SM, but MFV as well, if the sign of the
asymmetry is opposite to the SM or if it is same-sign but much enhanced.  We
also comment on the CP asymmetry in semileptonic $B_s$ decay, and update the
range of the angle $\beta_s$ in the SM: $0.026 < \sin2\beta_s < 0.048$.

\end{abstract}

\maketitle

\section{Introduction}
The CP asymmetry in semileptonic $B$ decays,
\beq\label{defasl}
A_{\rm SL}={\Gamma[\Bbar^0_{\rm phys}(t)\to\ell^+X]
-\Gamma[B^0_{\rm phys}(t)\to\ell^-X]\over
\Gamma[\Bbar^0_{\rm phys}(t)\to\ell^+X]
+\Gamma[B^0_{\rm phys}(t)\to\ell^-X]},
\eeq
depends on the relative phase between the absorptive and dispersive parts
of the $B^0-\Bbar^0$ mixing amplitude~\cite{old},
\beq\label{asldep}
A_{\rm SL}=\Im(\Gamma_{12}/M_{12}).
\eeq
Within the Standard Model (SM), the asymmetry is very small because of two 
suppression factors. First, the magnitude of the ratio is small, 
$|\Gamma_{12}/M_{12}|={\cal O}(m_b^2/m_t^2)\ll1$. Second, the phase is small,
$\arg(\Gamma_{12}/M_{12})={\cal O}(m_c^2/m_b^2)\ll1$. Since new physics
contributions to $\Gamma_{12}$ are small, and since $|M_{12}|$
is known from the measured value of the mass difference between the neutral
$B$ mesons, $\Delta m_B$, the first suppression factor should be valid model 
independently. In contrast, the second suppression factor could easily be 
avoided if new physics modifies the phase of $M_{12}$. This situation, where
new physics could enhance $A_{\rm SL}$ by a factor of ${\cal O}(10)$ makes 
this asymmetry a sensitive probe of new physics.

Recently, the search for CP violation in semileptonic $B$ decays achieved
a much improved sensitivity:
\beq\label{aslexpa}
A_{\rm SL}=\cases{(0.4\pm5.7)\times10^{-2}&OPAL \cite{Abbiendi:2000av},\cr
(1.4\pm4.2)\times10^{-2}&CLEO \cite{Jaffe:2001hz},\cr
(-1.2\pm2.8)\times10^{-2}&ALEPH \cite{Barate:2001uk},\cr
(0.5\pm1.8)\times10^{-2}&BABAR \cite{Aubert:2002mn}.\cr}
\eeq
The present world average is \cite{Nir:2001ge}
\beq
\label{aslexp}
A_{\rm SL}^{\rm exp}=(0.2\pm1.4)\times10^{-2}.
\eeq
With its experimental accuracy of order one percent, the result (\ref{aslexp})
puts for the first time meaningful constraints on new physics contributions to
$B^0-\Bbar^0$ mixing. It is our goal in this work to study these implications.

The plan of this paper is as follows. In section II we update the Standard
Model prediction for $A_{\rm SL}$, taking into account the recent measurements
of the CP asymmetry in $B\to\psi K_S$ decays. In section III we explain how
generic new physics can affect $A_{\rm SL}$. In section IV we  investigate the
effects of models of minimal flavor violation.  In each of sections II, III and
IV we first derive analytic expressions for the asymmetry and then carry out a
numerical investigation using the methods of ref.~\cite{Hocker:2001xe}. We give
our conclusions in section V.

\section{\boldmath $A_{\rm SL}$ in the Standard Model}
\subsection{Analytical Expressions}

Using the Standard Model expressions for $\Gamma_{12}$
\cite{Beneke:1996gn,Beneke:1998sy,Cahn:1999gx,Dighe:2001gc} and $M_{12}$, one
obtains
\beq\label{aslsml}
\ASL^0=-\kappa\ \Im\left({V_{cb}V_{cd}^*\over V_{tb}V_{td}^*}\right),
\eeq
where
\beq\label{defksl}
\kappa=4\pi{m_b^2\over m_W^2}
{K_1+K_2\over\bar\eta_B\, S_0(m_t^2/m_W^2)}\, z, \qquad
z\equiv m_c^2/m_b^2.
\eeq
This is the leading order result in the limit $m_b \gg \Lambda_{\rm QCD}$ and 
$z \ll 1$. Here $K_1$ and $K_2$ are Wilson coefficients, $\bar\eta_B$ is a QCD
correction factor and $S_0$ is the Inami-Lim function for the box diagram. The
CKM dependence can be expressed in terms of the $\rhobar$ and $\etabar$
parameters,
\beq\label{leadwol}
\Im\left({V_{cb}V_{cd}^*\over V_{tb}V_{td}^*}\right)=
{\etabar\over(1-\rhobar)^2+\etabar^2}\,.
\eeq

Eq.~(\ref{aslsml}) has three types of corrections characterized by small 
parameters: 
\beq\label{smhior}
A_{\rm SL}^{\rm SM}=A_{\rm SL}^0\left(1+a_{\rm SL}^z+a_{\rm SL}^{1/m_b}+
a_{\rm SL}^{\alpha_s}\right).
\eeq
The $m_c^2/m_b^2$ corrections are given by
\beqa\label{aslz}
a_{\rm SL}^z&=&-z{K_2\over K_1+K_2}+{z^2\over3}{K_2-K_1\over K_1+K_2}
-{2\over3}z(3-2z){K_2-K_1\over K_1+K_2}{\langle Q_S\rangle\over\langle Q\rangle}
\no\\
&-& {2(1-\rhobar)\over(1-\rhobar)^2+\etabar^2}\Bigg\{
{(1+2z)[2(1-z)^2-\sqrt{1-4z}]-1\over3z}{K_2-K_1\over K_1+K_2}
{\langle Q_S\rangle\over\langle Q\rangle}\\
&+& {(K_1+\frac{K_2}{2})[\sqrt{1-4z}-1+4z-2z^2]+z(1-z)^2(K_2-K_1)
-z\sqrt{1-4z}(K_1+2K_2)\over3z(K_1+K_2)}\Bigg\}.\no
\eeqa 
The matrix elements $\langle Q\rangle$ and $\langle Q_S\rangle$
can be parameterized as follows:
\beqa\label{defB}
\langle\Bbar|(\bar b_id_i)_{V-A}(\bar b_jd_j)_{V-A}|B\rangle&=&
{8\over3}f_B^2m_B^2B,\no\\
\langle\Bbar|(\bar b_id_i)_{S-P}(\bar b_jd_j)_{S-P}|B\rangle&=&
-{5\over3}f_B^2m_B^2{m_B^2\over(m_b+m_d)^2}B_S\equiv
-{5\over3}f_B^2m_B^2B_S^\prime.
\eeqa
In particular, we have ${\langle Q_S\rangle/\langle Q\rangle}=
-(5/8)(B_S^\prime/B)$. Some insight into the effect of $a_{\rm SL}^z$ can be 
gained by evaluating (\ref{aslz}) to ${\cal O}(z^2)$:
\beqa\label{immgz}
a_{\rm SL}^z
&=&z\left({5\over4}{K_2-K_1\over K_1+K_2}{B^\prime_S\over B}-
{K_2\over K_1+K_2}\right)\no\\
&+& z^2\left[{K_2-K_1\over K_2+K_1}\left({1\over3}-{5\over6}{B^\prime_S\over B}
\right)+2{1-\rhobar\over(1-\rhobar)^2+\etabar^2}
{K_2-K_1\over K_1+K_2}\left({5\over2}{B_S^\prime\over B}-1\right)\right].
\eeqa
Note that the terms with CKM dependence that is different from the leading
result appear only at ${\cal O}(z^2)$ and are therefore very small.  

The $1/m_b$ corrections are given by
\beqa\label{gamoom}
a_{\rm SL}^{1/m_b}&=&2z{-2K_1\left[\langle R_1\rangle
-2\langle R_3\rangle\right]+K_2\left[\langle R_2\rangle+4\langle R_3\rangle
+2\langle R_4\rangle\right]\over (K_1+K_2)\langle Q\rangle}\\
&+&4z^2\left[{1\over3}-{2(1-\rhobar)\over(1-\rhobar)^2+\etabar^2}\right]
{K_1\left[2\langle R_1\rangle-\langle R_2\rangle-6\langle R_3\rangle\right]
-K_2\left[\langle R_2\rangle+6\langle R_3\rangle+2\langle R_4\rangle\right]
\over (K_1+K_2)\langle Q\rangle}.\no
\eeqa
The matrix elements $\langle R_i\rangle$ have the following values within the
vacuum insertion approximation:
\beqa\label{matele}
\langle R_1\rangle&=&{7\over3}{m_d\over m_b}f_B^2m_B^2,\no\\
\langle R_2\rangle&=&-{2\over3}{m_B^2-m_b^2\over m_b^2}f_B^2m_B^2,\no\\
\langle R_3\rangle&=&{7\over6}{m_B^2-m_b^2\over m_b^2}f_B^2m_B^2,\no\\
\langle R_4\rangle&=&-{m_B^2-m_b^2\over m_b^2}f_B^2m_B^2.
\eeqa
Note that ${\langle R_1\rangle / \langle Q\rangle}={\cal O}\left(
{m_d / m_b}\right)$ and ${\langle R_{2,3,4}\rangle / \langle Q\rangle}=
{\cal O}\left({\Lambda_{\rm QCD} / m_b}\right)$. We can therefore
safely neglect terms of order $z\langle R_1\rangle/\langle Q\rangle$ and
$z^2\langle R_{2,3,4}\rangle/\langle Q\rangle$. In this approximation we obtain
\beq\label{asloom}
a_{\rm SL}^{1/m_b}=z\, {7K_1+3K_2\over 2(K_1+K_2)}\,
{1\over B}\, {m_B^2-m_b^2\over m_b^2}.
\eeq

The ${\cal O}(\alpha_s)$ corrections have not been fully calculated. They can 
be divided into penguin corrections and NLO corrections. The penguin terms give
\beqa\label{gampen}
a_{\rm SL}^{\rm peng}&=&{K_1^\prime+K_2^\prime-K_3^\prime\over K_1+K_2}\no\\
&+&z\left\{{2K_3^\prime-K_2^\prime\over K_1+K_2}
+2{K_1^\prime-K_2^\prime\over K_1+K_2}{\langle Q_S\rangle\over\langle Q\rangle}
\right.\no\\
&+&\left.4{K_1^\prime\left[2\langle R_3\rangle-\langle R_1\rangle\right]
+6K_2^\prime\left[\langle R_2\rangle+4\langle R_3\rangle+2\langle R_4\rangle
\right]+4K_3^\prime\langle R_2\rangle\over(K_1+K_2)\langle Q\rangle}\right\}.
\eeqa
While $K_{1,2}$ are combinations of the Wilson coefficients $C_{1,2}$,
the $K^\prime_{1,2,3}$ depend also on $C_{3,4,5,6}$ which are suppressed
by $\alpha_s/\pi$:
\beqa\label{wilcoe}
K_1&=&3C_1^2+2C_1C_2,\no\\
K_2&=&C_2^2,\no\\
K_1^\prime&=&2(3C_1C_3+C_1C_4+C_2C_3),\no\\
K_2^\prime&=&2C_2C_4,\no\\
K_3^\prime&=&2(3C_1C_5+C_1C_6+C_2C_5+C_2C_6).
\eeqa
Given that $K^\prime_i/K_j = {\cal O}(\alpha_s/\pi)$, we can safely neglect
terms of order $zK^\prime_i$.

The NLO corrections to $A_{\rm SL}$, that is, corrections to $\Gamma_{12}$ of 
${\cal O}(z\alpha_s)$, have not been calculated. The challenge lies in the
diagrams involving a charm quark, an up quark and a gluon in the intermediate
state, which are very sensitive to $m_c$, but have only been computed in the
$m_u=m_c$ limit~\cite{Beneke:1998sy}. Consequently, there is an ambiguity as to
which definition of $m_c$ is best suited to the evaluation of $A_{\rm SL}$.
This is the largest uncertainty at present in the Standard Model prediction for
this asymmetry.

To summarize, the Standard Model expression for the CP asymmetry in
semileptonic $B$ decays is given by
\beqa\label{immgacor}
A_{\rm SL}^{\rm SM}=&-&\kappa\,{\etabar\over(1-\rhobar)^2+\etabar^2}
\left[1+z\left({5\over4}{B^\prime_S\over B}\,{K_2-K_1\over K_1+K_2}
-{K_2\over K_1+K_2}\right)\right.\no\\*
&+& z^2\,{K_2-K_1\over K_2+K_1}\left({1\over3}
-{5\over6}{B^\prime_S\over B}\right)
+2z^2\,{1-\rhobar\over(1-\rhobar)^2+\etabar^2}\,
{K_2-K_1\over K_1+K_2}\left({5\over2}{B_S^\prime\over B}-1\right)\no\\*
&+&\left.z\,{7K_1+3K_2\over2(K_1+K_2)}\,{1\over B}{m_B^2-m_b^2\over m_b^2}
+{K_1^\prime+K_2^\prime-K_3^\prime\over K_1+K_2}
+{\cal O}\left({\alpha_s\over4\pi}\right)\right].
\eeqa

\subsection{Numerical Results}

There are a number of input parameters needed to evaluate eq.~(\ref{immgacor}).
The ones not discussed explicitly below ($m_t$, $|V_{cb}|$, $B_K$, etc.) are
taken from ref.~\cite{Hocker:2001jb}. The $K$'s are calculable in perturbation
theory and we use their values in the NDR scheme with
$\Lambda^{(5)}_{\overline{\rm MS}}=225\,$MeV (which is close to
$\alpha_s(m_Z)=0.118$), given in Table XIII of ref.~\cite{Buchalla:1996vs}. 
This gives the values shown in Table~\ref{tab:inputs}.  The uncertainty related
to these is tiny compared to the ones discussed next, and will be neglected. 
For the bag parameters we use the unquenched lattice QCD results with two light
flavors~\cite{Yamada:2001xp,Shoji}; besides the published values we also use
$B_{S_s}/B_{S_d} = 1.04(2)$~\cite{Shoji}.  These results are in good agreement
with~\cite{Becirevic:2001xt}.  We also use $f_B = 200 \pm 30\,$MeV
\cite{Ryan:2001ej}, which is consistent with $f_B \sqrt{(\bar\eta_B/\eta_B)
B(m_b)} \simeq 1.15 f_B \simeq 230 \pm 30\,$MeV used in \cite{Hocker:2001jb}. 
Here, and in the definition of $\kappa$ in eq.~(\ref{defksl}), $\bar\eta_B$ is
related to the scale independent $\eta_B$ via $\bar\eta_B = \eta_B
[\alpha_s(m_b)]^{-6/23} \left(1+{\alpha_s\over4\pi} {5165\over
3174}\right)$~\cite{Buchalla:1996vs}.

\begin{table}[ht]
\begin{center}
\begin{tabular}{cc} \hline \hline       
~~Input Parameter~~	&  Value 		\\ \hline
$K_1$			&  $-0.295$		\\
$K_2$			&  ~~$1.162$		\\
$K'_1$			&  ~~$0.027$		\\
$K'_2$			&  $-0.076$		\\
$K'_3$			&  $-0.064$		\\ 
$\alpha_s(m_b)$		&  $0.22$		\\
$B(m_b)$		&  $0.87 \pm 0.04 \pm 0.07$	\\
$B_S(m_b)$		&  $0.83 \pm 0.03 \pm 0.07$	\\
$f_B$			&  $(200 \pm 30)\,$MeV		\\
$\eta_B$		&  $0.55 \pm 0.01_{\rm theo}$	\\
$m_b^{\rm pole}$	&  $(4.8 \pm 0.1_{\rm theo})\,$GeV \\
$z$			&  $0.085 \pm 0.01_{\rm theo}$	\\
\hline\hline
\end{tabular}
\end{center}
\caption{Inputs for the fits, other than those listed in \cite{Hocker:2001jb}. 
When no error is stated, the quantity is held constant.  Errors with ``theo''
subscripts are treated as ranges, and as Gaussians otherwise.}
\label{tab:inputs}
\end{table}

At present, the biggest uncertainty in evaluating eq.~(\ref{immgacor}) comes
from not knowing the NLO [${\cal O}(\alpha_s)$] corrections.  In particular, it
results in an ambiguity in the quark mass definitions used.  This is not
relevant for $z$, since $z$ is only scheme dependent at NNLO [${\cal
O}(\alpha_s^2)$].  We use $z=0.085\pm0.01$, which takes into account the
correlation between $m_b$ and $m_c$.  In $\aSL^{1/m_b}$, in eq.~(\ref{asloom}),
we use the pole mass~\cite{Dighe:2001gc}, $m_b^{\rm pole} = (4.8 \pm
0.1)\,$GeV.  In the definition of $B_S'$, in eq.~(\ref{defB}), it is the
$\ov{\rm MS}$ mass which enters, and we use the one-loop relation,
$\ov{m}_b(m_b) = m_b^{\rm pole} \left( 1 - {4\alpha_s\over3\pi} \right)$.  The
$m_b^2$ factor which enters the definition of $\kappa$ in eq.~(\ref{defksl}) is
a large source of uncertainty that will only be reduced when the NLO correction
is known, and we choose to use $\ov{m}_b(m_b)$.  While this choice is somewhat
ad hoc, it is motivated by the fact that it is a good approximation in the case
when $\Gamma_{12}$ is known to NLO precision, i.e., in the $m_u = m_c$ limit.
This is also a conservative choice for constraining new physics in the rest of
this paper (since it reduces the SM expectation compared to using $m_b^{\rm
pole}$).

\begin{figure}[t]
\centerline{\epsfxsize=0.5\textwidth \epsfbox{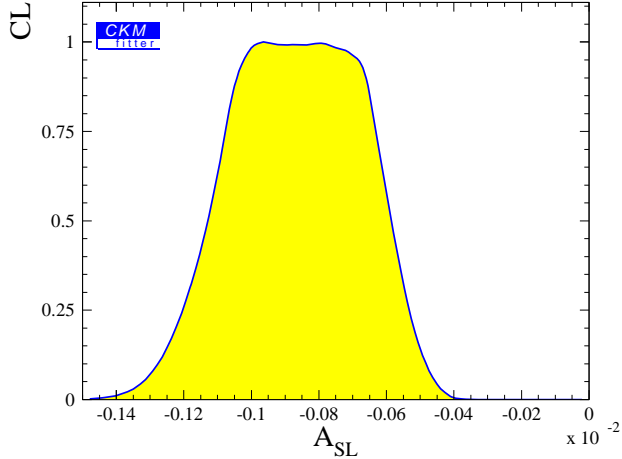}}
\caption{Confidence levels of $\ASL$ in the Standard Model.}
\label{aslsmrange}
\end{figure}

In Fig.~\ref{aslsmrange} we show the Standard Model prediction for $\ASL$,
using the $R$fit approach in the \verb"CkmFitter" package~\cite{Hocker:2001xe},
where theoretical errors are considered as allowed ranges
\cite{Harrison:1998yr,Hocker:2001xe}.\footnote{In this approach, one determines
the confidence level (CL) for a particular set of parameter values (e.g.,
$\rhobar-\etabar$, $\ASL$, etc.) to be consistent with both the measurements
and the theoretical inputs, the latter being let free to vary within their
allowed ranges.  For example, the curve in Fig.~\ref{aslsmrange} gives the CL
that a certain value of $\ASL$ is consistent with the theory.  But the CL of
$\ASL$ falling within a range is not defined, since that would require all
input parameters to be viewed as distributions with probabilistic
interpretations.}  With the above input parameters, the range of $\ASL$ values
with greater than 10\% CL is
\beq\label{asl90}
-1.28\times10^{-3} < \ASL < -0.48\times10^{-3}\,.
\eeq
The range with greater than 32\% CL (that would correspond to a ``$1\sigma$
range" if the theoretical errors were negligible) is only slightly smaller,
$-1.18\times10^{-3} < \ASL < -0.55\times10^{-3}$. This indicates that the
uncertainty in eq.~(\ref{asl90}) is dominated by theoretical errors, resulting
in the plateau in Fig.~\ref{aslsmrange} with a confidence level near unity. 
The uncertainty will decrease when the constraints on $\etabar$ improve, the
NLO correction to $\Gamma_{12}$ is computed, and the $b$ quark mass is more
precisely determined.  We did not assign an error to the assumption of local
quark-hadron duality which enters the OPE calculation of the nonleptonic $B$
decay rates determining $\Gamma_{12}$.  We may gain confidence that the errors
related to this are small if future lattice calculations can account for the
$b$ hadron lifetimes and, especially, for the $B_s$ lifetime difference.

\begin{figure}[htb]
\centerline{\epsfxsize=0.5\textwidth \epsfbox{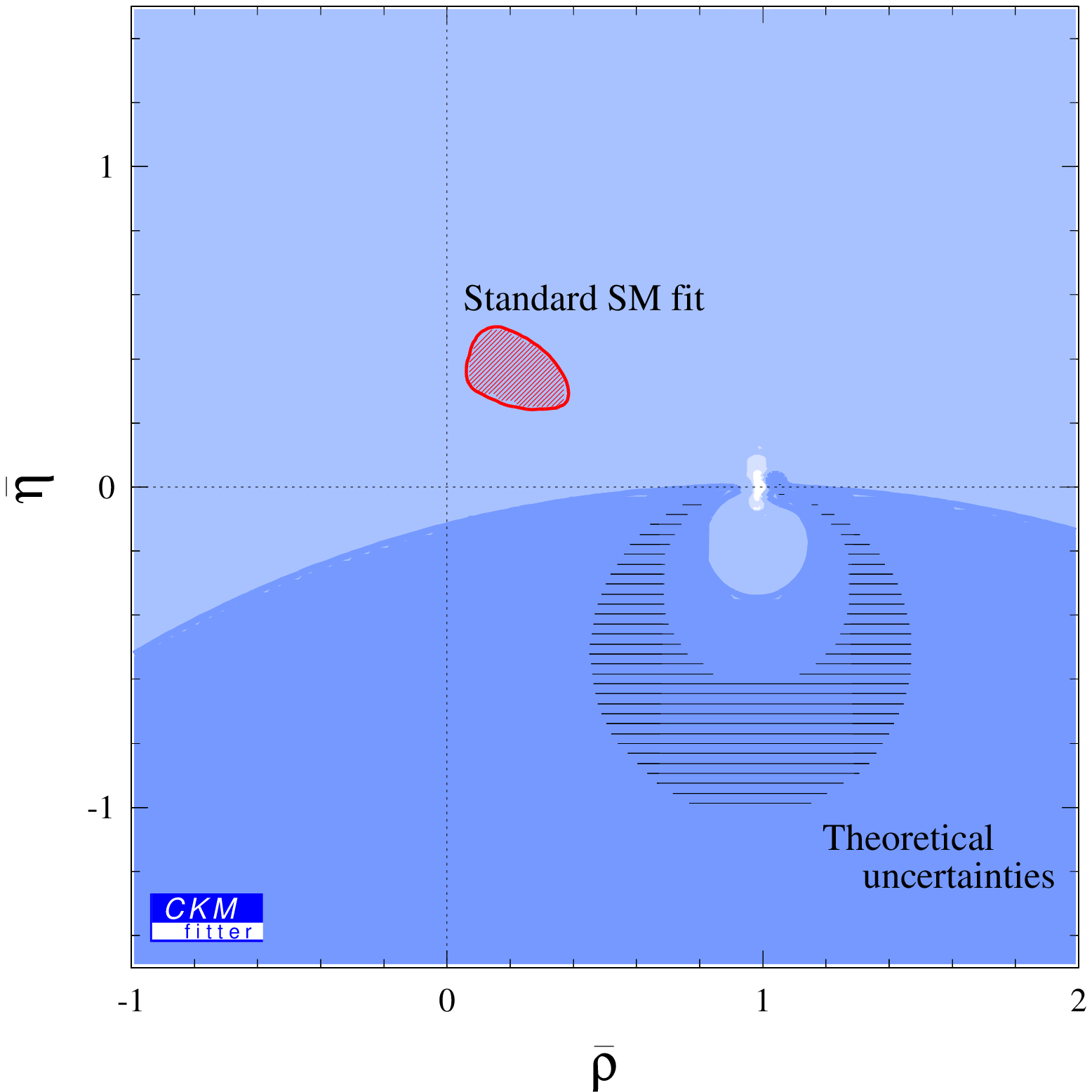} \hfil
  \epsfxsize=0.5\textwidth\epsfbox{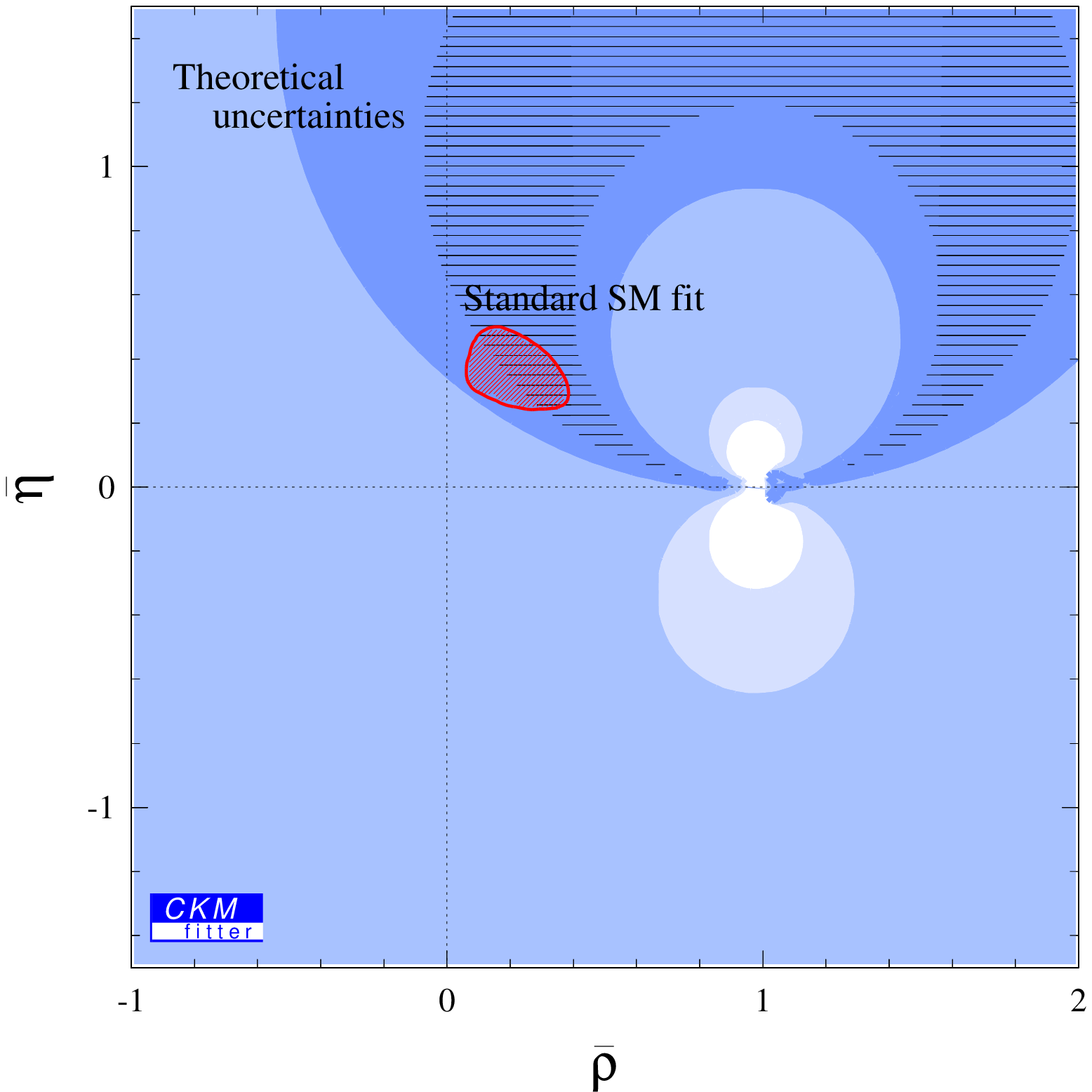}}
\caption{The constraint on the $\rhobar-\etabar$ plane from the present data
on $\ASL$ in eq.~(\ref{aslexp}) (left); and the constraint that would follow
from a hypothetical value $\ASL = (-1 \pm 3)\times 10^{-3}$ (right). The central
value of the latter was chosen to be consistent with the SM, and the error may 
be achievable by 2005.  The dark-, medium-, and light-shaded regions have 
confidence levels above 90\%, 32\%, and 10\%, respectively.  The white regions
contain points with at most 10\%~CL.}
\label{aslsmrhoeta}
\end{figure}

At leading order, corresponding to $\ASL^0$ in eq.~(\ref{aslsml}), a constraint
$X_- < \ASL < X_+$ with $X_+ > 0$ and $X_- < 0$ excludes the interior of two
circles, one with radius $R_+$ around the point $(\rhobar, \etabar) = (1,
-R_+)$, and another with radius $R_-$ around the point $(\rhobar, \etabar) =
(1, R_-)$.  With the above central values of the input parameters, $R_\pm =
\kappa/(2|X_\pm|) \simeq 6.7\times 10^{-4}/|X_\pm|$.  If $X_+ < 0$ [$X_- > 0$],
then the corresponding excluded region is the outside of a circle of radius
$R_+$ [$R_-$] around the point $(1, R_+)$ [$(1, -R_-)$].  Including the terms
in $\aSL^z$ distorts these circles by making them slightly larger for
$\rhobar<1$ than for $\rhobar>1$, as shown in Fig.~\ref{aslsmrhoeta}.

In Fig.~\ref{aslsmrhoeta}, the left plot shows the constraint that the present
data on $\ASL$, eq.~(\ref{aslexp}), provide on the $\rhobar-\etabar$ plane. The
right plot shows the constraint that would follow from a hypothetical future
value $\ASL = (-1 \pm 3)\times 10^{-3}$. We chose the central value to be
within the SM allowed range and the error to represent an experimental accuracy
that may be achievable with 500\,fb$^{-1}$ data expected by 2005 at the $B$
factories, using a simple statistical scaling of the BABAR
result~\cite{Aubert:2002mn}, which is based on 20\,fb$^{-1}$ data and is the
most precise one in eq.~(\ref{aslexpa}).  The dark shaded regions contain the
points with confidence levels above 90\%, and include the best fit points.  The
dark and medium shaded regions together contain the ``one sigma" allowed
regions with CL above 32\%. The dark-, medium-, and light-shaded regions 
together contain all points with CL higher than 10\%.  Thus, the white regions
have at most 10\% CL.  The small diagonally hatched areas show the SM allowed
region (points with greater than 10\% CL).  The horizontally stripped regions
contain the points with CL above 10\% for a hypothetical ``perfect" measurement
of $A_{\rm SL} = +0.002$ (left plot) and $A_{\rm SL} = -0.001$ (right plot). 
These illustrate the significance of the present theoretical errors in
interpreting the experimental results, and the importance of reducing them by
determining $m_b$ and $z$ more precisely and especially by completing the NLO
calculation of $\Gamma_{12}$.

One sees that within the next few years $\ASL$ will not be a useful constraint
on $(\rhobar, \etabar)$ in the context of the Standard Model if the
experimental result remains consistent with the SM prediction.  However, as we
discuss it next, it is a sensitive probe of new physics, and already provides
interesting constraints on certain models.

\section{\boldmath $A_{\rm SL}$ with New Physics}
\subsection{Analytical Expressions}

We investigate $\ASL$ in models of new physics 
\cite{Sanda:1997rh,Randall:1998te,Cahn:1999gx,Barenboim:1999in}
with the following two features:

(i) The $3\times3$ CKM matrix is unitary.

(ii) Tree level processes are dominated by the SM.

The second feature means that $\Gamma_{12}=\Gamma_{12}^{\rm SM}$, and the new 
physics effects modify only $M_{12}$. Then, quite generally, these effects can
be parameterized by two new parameters, $\theta_d$ and $r_d$, defined through
\beq\label{motnp}
M_{12}=r_d^2\, e^{2i\theta_d} M_{12}^{\rm SM}.
\eeq
This modification involves well-known consequences for the mass difference
between the neutral $B$ mesons,
\beq\label{delmnp}
\Delta m_B=r_d^2(\Delta m_B)^{\rm SM},
\eeq
\OMIT{
where
\beq\label{dmbsm}
(\Delta m_B)^{\rm SM}={G_F^2\over6\pi^2}m_W^2m_B B f_B^2\bar\eta_B
S_0(m_t^2/m_W^2)|V_{tb}V_{td}|^2,
\eeq
}%
and for the CP asymmetry in charmonium-containing $b\to c\bar cs$ decays, which
is denoted throughout this paper by $a_{\psi K}$,
\beq\label{apknp}
a_{\psi K}=\sin(2\beta+2\theta_d).
\eeq
For the CP asymmetry in semileptonic decays, the modification from the Standard
Model value depends on both $r_d^2$ and $2\theta_d$:
\beq\label{aslnp}
A_{\rm SL}=-\Re\left({\Gamma_{12}\over M_{12}}\right)^{\rm SM} 
{\sin2\theta_d\over r_d^2}+\Im\left({\Gamma_{12}\over M_{12}}\right)^{\rm SM}
{\cos2\theta_d\over r_d^2}.
\eeq
The first term has been previously investigated. Since  ${\rm Re} ({\Gamma_{12}
/ M_{12}})^{\rm SM}$ is larger than $A_{\rm SL}^{\rm SM}$ by a factor of order
$m_b^2/m_c^2$, it could give an asymmetry that is an order of magnitude larger
than the Standard Model prediction. This would happen if the new physics
contribution to $M_{12}$ has a large new phase  ($\sin2\theta_d\not\ll1$).  The
second term is suppressed by $m_c^2/m_b^2$ compared to the first one, however,
one might expect it to be enhanced in the region $\sin2\theta_d \approx 0$ and
$r_d^2 \ll 1$, corresponding to cancelling contributions to $\Delta m_B$ from
the Standard Model and from new physics. However, we find that this term plays
a numerically negligible role as long as the error of $\ASL$ is much larger
than the SM expectation.

Our purpose is to find the constraints on $r_d^2$ and $2\theta_d$ from the
present measurements of $A_{\rm SL}$. We need therefore to evaluate
$(\Gamma_{12}/M_{12})^{\rm SM}$. In ref.~\cite{Dighe:2001gc} one can find the
$m_c^2/m_b^2$, $1/m_b$, penguin and NLO QCD corrections. Given the present
experimental accuracy, it is sufficient for our purposes to include  only
corrections of order 10\%, that is, ${\cal O}(z)$ and ${\cal O}(1/m_b)$
corrections:
\beqa\label{gotsman}
\left({\Gamma_{12}\over M_{12}}\right)^{\rm SM}
=&-&{4\pi m_b^2\over3m_W^2 \bar\eta_B S_0(m_t^2/m_W^2)}\,
\Bigg[{5\over8}(K_2-K_1){B_S^\prime\over B}+\left(K_1+{K_2\over2}\right) \no\\
&+&\left({K_1\over2}-K_2\right){m_B^2-m_b^2\over m_b^2}{1\over B}
-3z(K_1+K_2){1-\rhobar-i\etabar\over(1-\rhobar)^2+\etabar^2}\Bigg] .
\eeqa

There are four parameters related to flavor violation that are relevant to our
discussion here: the CKM parameters $\rhobar$ and $\etabar$ and the new physics
parameters of eq.~(\ref{motnp}) $r_d^2$ and $2\theta_d$. There are also four
critical constraints: $|V_{ub}|$ from charmless semileptonic $B$ decays (these
are tree level processes and therefore, by assumption, unaffected by the new
physics), $\Delta m_B$ of eq.~(\ref{delmnp}), $a_{\psi K}$ of
eq.~(\ref{apknp}), and $A_{\rm SL}$ of eq.~(\ref{aslnp}).  In the next
subsection we study the consistency of these four measurements and their
combined constraints on the model parameters. Note that we cannot use the
measured value of $\epsilon_K$ and the lower bound on $\Delta m_{B_s}$ since
they may involve additional parameters.

\subsection{Numerical Results}

\begin{figure}[t]
\centerline{\epsfxsize=0.5\textwidth\epsfbox{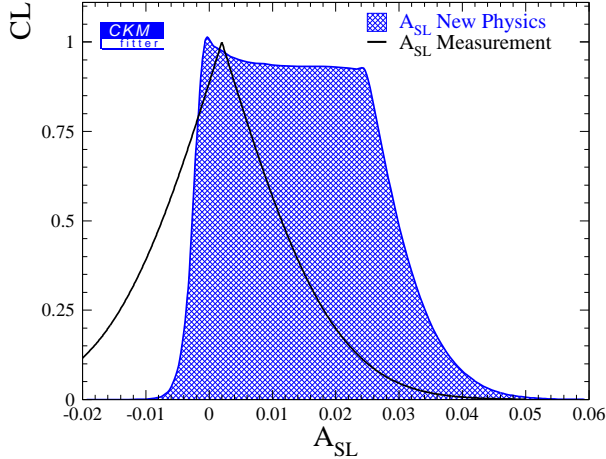}}
\caption{The filled area shows the confidence levels of $\ASL$ for arbitrary
$r_d$ and $\theta_d$ using the constraints from $|V_{ub}|$, $\Delta m_B$, and
$a_{\psi K}$.  The curve represents the measured value of $\ASL$ in
eq.~(\ref{aslexp}), and indicates that some new physics models are being
excluded.}
\label{fig:rdthetadrange}
\end{figure}

Fig.~\ref{fig:rdthetadrange} shows the allowed range of $\ASL$ for arbitrary
$r_d$ and $\theta_d$ using the constraints from $|V_{ub}|$, $\Delta m_B$, and
$a_{\psi K}$.  To evaluate eq.~(\ref{gotsman}), we used the $\ov{\rm MS}$ $b$
quark mass in the overall factor, as in sec.~II.B.  The black curve
superimposed in Fig.~\ref{fig:rdthetadrange} shows the confidence level
corresponding to the measured value of $\ASL$ in eq.~(\ref{aslexp}).  In the
presence of new physics that can be parameterized by $r_d$ and $\theta_d$, the
range of $\ASL$ with greater than 10\% CL is $-0.4\times 10^{-2} < \ASL <
3.9\times 10^{-2}$, whereas the measurement in eq.~(\ref{aslexp}) implies
$-2.1\times 10^{-2} < \ASL < 2.5\times 10^{-2}$ at the same CL.  Clearly, the
recent measurements of $\ASL$ are sensitive enough to probe new physics, and
already constrain the $r_d^2 - 2\theta_d$ parameter space.

It is interesting to note that if $\ASL$ is negative, as in the SM, then new
physics that can be parameterized by $r_d$ and $\theta_d$ can only enhance it
by a factor of few.  However, if new physics makes $\ASL$ positive, then it can
be enhanced by more than an order of magnitude. The reasons for this situation
are explained below.

\begin{figure}
\centerline{\epsfxsize=0.4\textwidth\epsfbox{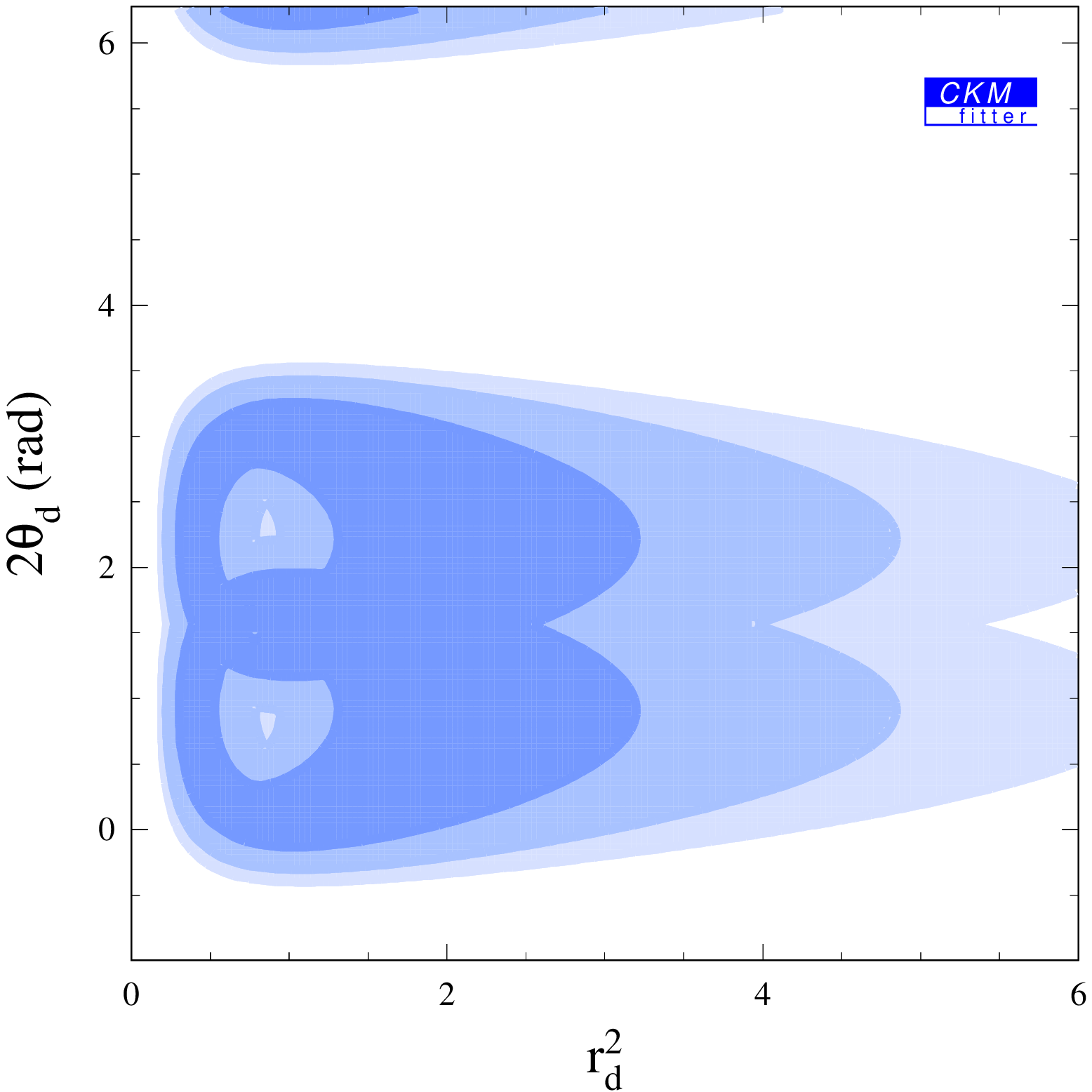} \ \ \
  \epsfxsize=0.4\textwidth\epsfbox{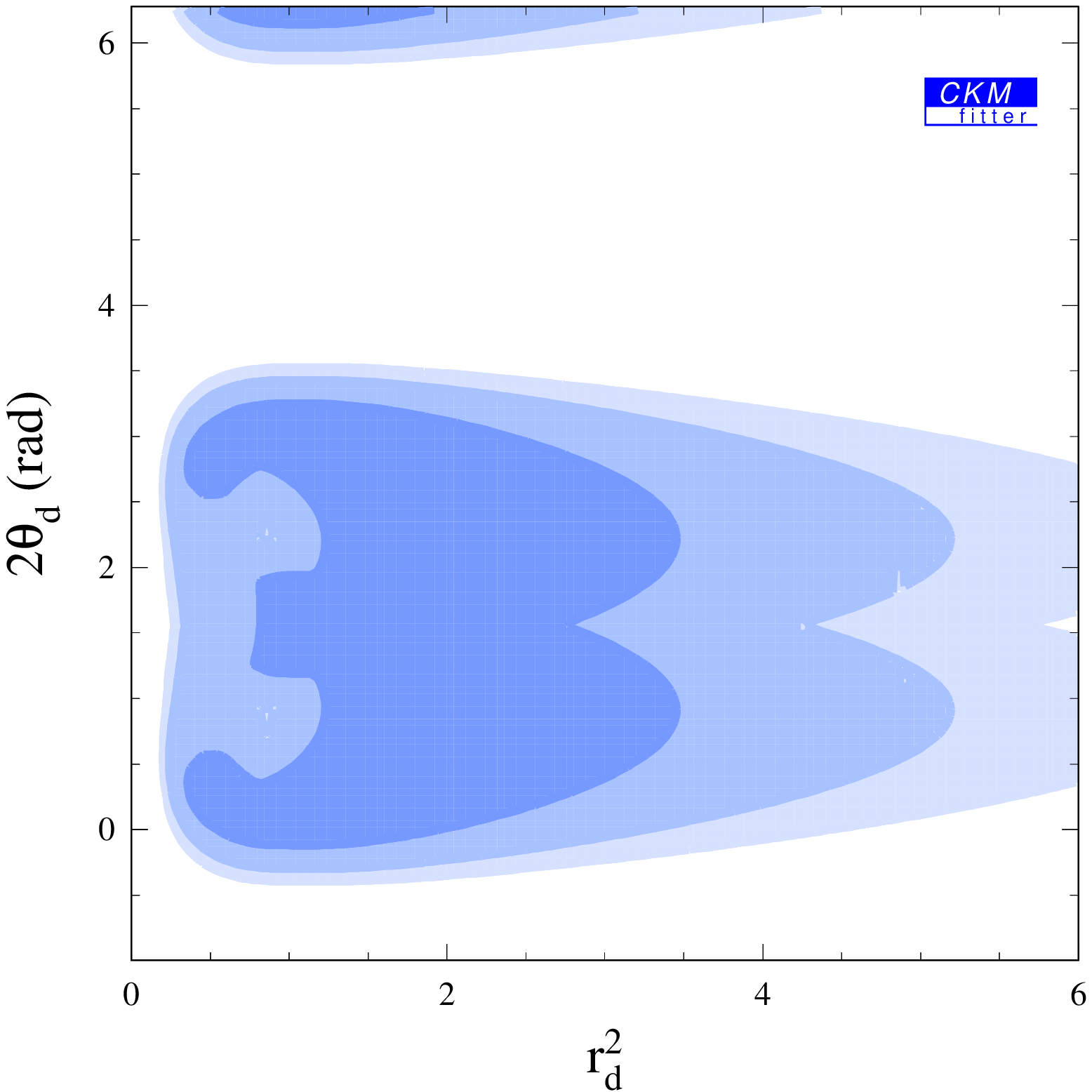} }
\centerline{\epsfxsize=0.4\textwidth\epsfbox{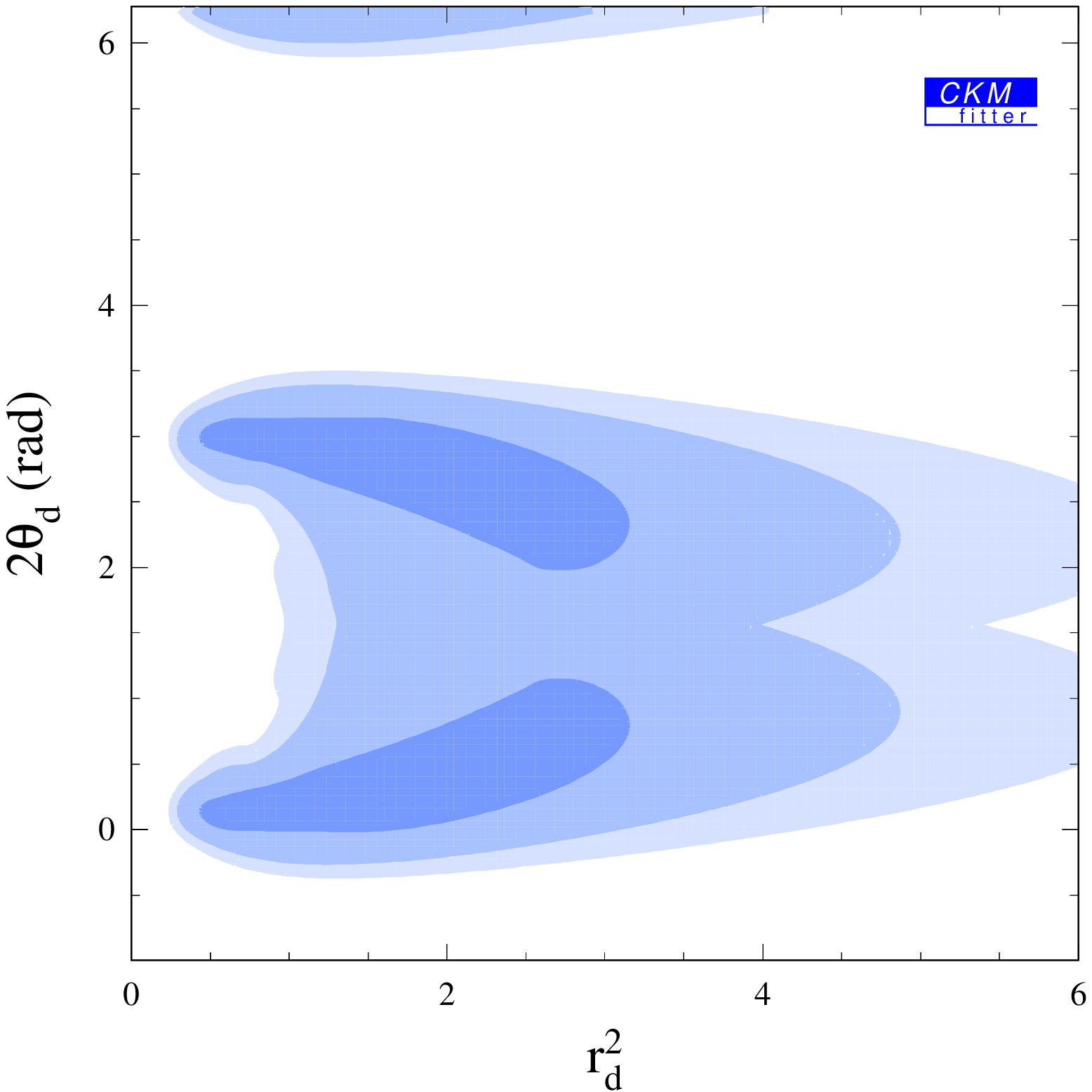} \ \ \ 
  \epsfxsize=0.4\textwidth\epsfbox{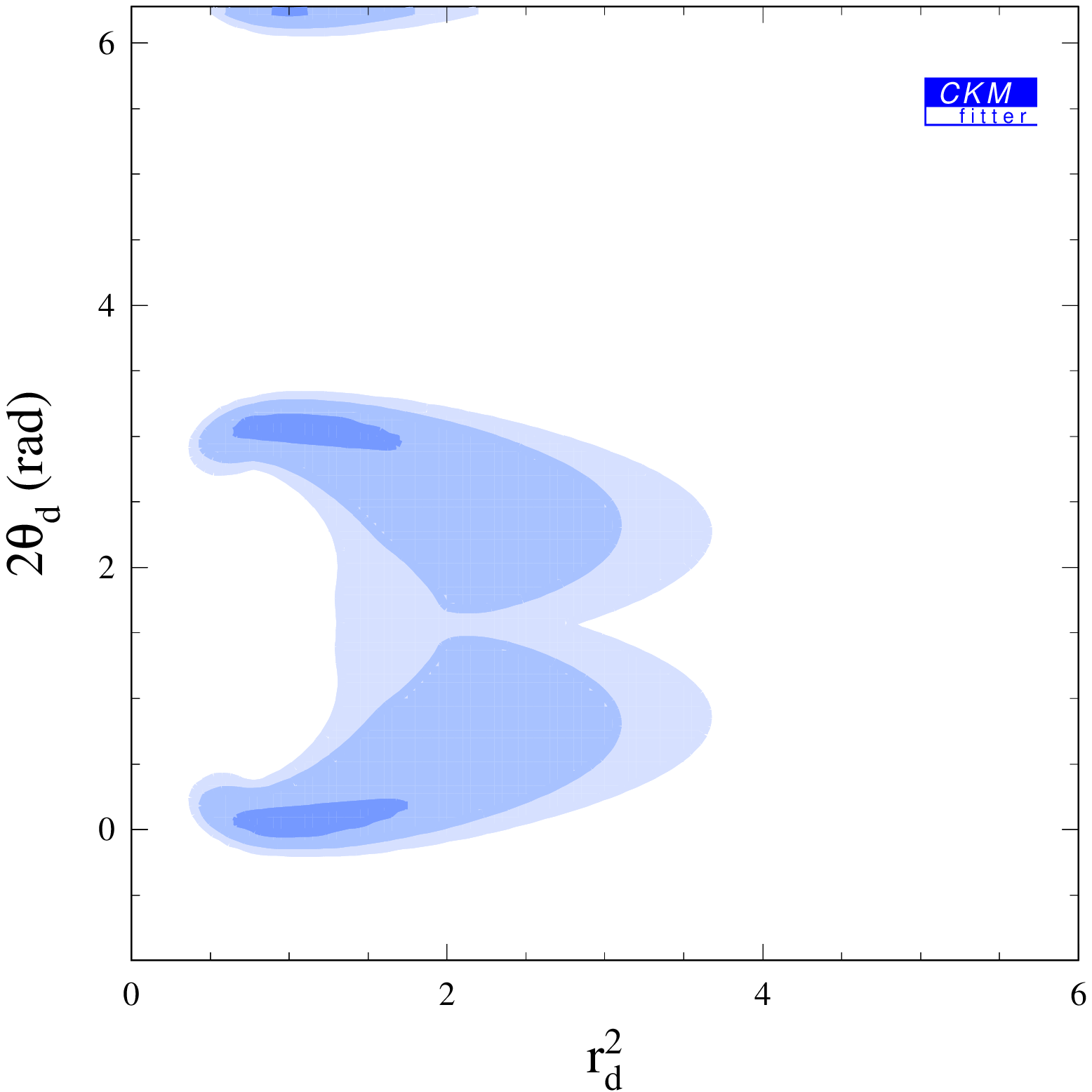}}
\caption{The top left plot shows the confidence levels of $r_d^2$ and
$2\theta_d$ from the measurements of $|V_{ub}|$, $\Delta m_B$, and $a_{\psi
K}$; the top right plot also includes $\ASL$ as a constraint.  The bottom left
plot shows the confidence levels corresponding to a  hypothetical future value
$\ASL = (-1 \pm 3)\times 10^{-3}$, while the bottom right plot assumes in
addition reduced errors in $|V_{ub}|$ (10\%), $f_B$ (10\,MeV) and $a_{\psi K}$
(0.04).  The shadings mean the same as in Fig.~\ref{aslsmrhoeta}.}
\label{fig:aslrdthetad}
\end{figure}

The top left plot in Fig.~\ref{fig:aslrdthetad} shows the confidence levels of
$r_d^2$ and $2\theta_d$ from the measurements of $|V_{ub}|$, $\Delta m_B$, and
$a_{\psi K}$.  The ranges with greater than 10\% CL are
\beq\label{rt90}
0.2 < r_d^2 < 6.4\,, \qquad -0.4 < 2\theta_d < 3.6 \,.
\eeq
The errors are again dominated by theoretical ranges, except for the upper
bound on $r_d^2$. Consequently, the ranges with greater than 32\% CL, $0.2 <
r_d^2 < 5.2$ and $-0.3 < 2\theta_d < 3.5$, only differ significantly from
eq.~(\ref{rt90}) in $(r_d^2)_{\rm max}$.  This plot makes it clear why new
physics can  enhance much more a positive $\ASL$ value than a negative one.
First, since $a_{\psi K}>0$ is  firmly established, the maximal allowed
magnitude of $\sin2\theta_d$ is larger for positive values than it is for
negative ones~\cite{Eyal:1999ii}. Second, the correlations between the three
constraints are such that while maximal positive $\sin2\theta_d$ and minimal
$r_d^2$ are allowed simultaneously, this  is not the case for negative
$\sin2\theta_d$.  The top right plot shows the present constraints on the
$r_d^2 - 2\theta_d$ plane, including also the measurement of $\ASL$.  The
measurement of $\ASL$ has an noticeable impact: in particular, the lowest
values of $r_d^2$ (around 0.2) are now disfavored, and almost the entire $r_d^2
\lsim 0.5$ region is no longer among the best-fit points.  The bottom left plot
show the constraints that would follow from a hypothetical future value, $\ASL
= (-1 \pm 3)\times 10^{-3}$.  Such a measurement would be able to exclude a
large part of the $r_d^2<1$ parameter space (i.e., cancelling contributions to
$\Delta m_B$ from the SM and new physics), except if $2\theta_d$ were near $0$
or $\pi$.  This plot is rather insensitive to the expected reduction of the
error of $a_{\psi K}$ by itself, while additional improvements in $|V_{ub}|$
and $\Delta m_B$  will make a significant difference.  This is shown in the
bottom right plot, where reduced errors about the present central values in
$|V_{ub}|$ (10\%), $f_B$ (10\,MeV) and $a_{\psi K}$ (0.04) are also assumed.

\subsection{\boldmath The $B_s$ system}

\begin{figure}
\centerline{\epsfxsize=0.5\textwidth\epsfbox{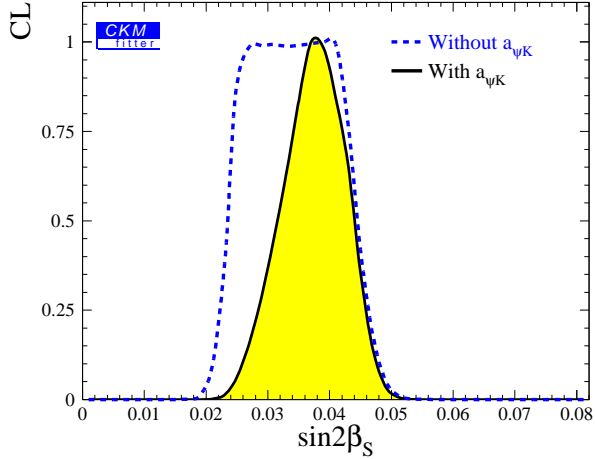}}
\caption{Confidence levels of $\sin 2\beta_s$ in the Standard Model.}
\label{fig:sin2bs}
\end{figure}

Similar analyses will become possible in the future for the $B_s$ system.
Within the SM, the semileptonic asymmetry in $B_s$ decays is suppressed even
more strongly than the asymmetry in $B_d$ decays. (For the sake of clarity, we
now call the respective asymmetries $A_{\rm SL}^s$ and $A_{\rm SL}^d$.) While
the $m_c^2/m_t^2$ suppression factor in eq. (\ref{defksl}) is common to both,
the CKM dependence is different. For $A_{\rm SL}^d$ it is given by 
eq.~(\ref{leadwol}), which is a factor of order unity. In contrast, for $A_{\rm
SL}^s$ it is given by $\Im\left({V_{cb}V_{cs}^*\over V_{tb}V_{ts}^*} \right)
\approx -\frac12 \sin2\beta_s$. In Fig.~\ref{fig:sin2bs} we give the CL for
$\sin2\beta_s$ within the SM. We learn that the range of $\sin2\beta_s$ with
greater than 10\% CL is
\begin{equation}
0.026 < \sin2\beta_s < 0.048 \,,
\end{equation}
Consequently, $A_{\rm SL}^s$ is unobservably small in the SM. 

Defining $M_{12}^s\equiv r_s^2 e^{2i\theta_s} (M_{12}^s)^{\rm SM}$, the
present lower bound from the LEP/SLD/CDF combined amplitude fit
(corresponding to $\Delta m_{B_s} > 15.0\, \mbox{ps}^{-1}$) implies
\beq
r_s^2 > 0.6 \,,
\eeq
whereas there is no constraint on $\theta_s$ yet. These parameters can be
determined in the future in the following ways: (i) A measurement of $\Delta
m_{B_s}$ will constrain $r_s^2$. (ii) A measurement of the CP asymmetry in
$B_s\to \psi\phi$ or $\psi\eta^{(\prime)}$ will constrain
$\sin2(\beta_s+\theta_s)$. If the asymmetry is much larger than the SM range,
it practically determines $\sin2\theta_s$. (iii) A measurement of $A_{\rm
SL}^s$ will be proportional to $\sin2\theta_s/r_s^2$ and will give a
consistency check on the interpretation of the previous two measurements. The
values of the four parameters $r_d^2$, $2\theta_d$, $r_s^2$ and $2\theta_s$
provide an excellent probe of the flavor and CP structure of new physics.

\section{\boldmath $A_{\rm SL}$ with Minimal Flavor Violation}
\subsection{Analytical Expressions}

Minimal flavor violation (MFV) is the name given to a class of new physics
models that do not have any new operators beyond those present in the Standard
Model and in which all flavor changing transitions are governed by the CKM
matrix with no new phases beyond the CKM phase
\cite{Ciuchini:1998xy,Ali:1999we,Buras:2000dm,Buras:2000xq,Buras:2001af,Buras:2001pn,Bergmann:2001pm}. 
Examples include the constrained minimal supersymmetric Standard Model and the 
two Higgs doublet models of types I and II. In these models, the SM predictions
for some flavor changing processes remain unchanged, while other  are modified
but in a correlated way. The new physics contributions that are relevant to our
discussion depend on a single new  parameter, $F_{tt}$, that is real but could
have either sign.  These models can be viewed as special cases of sec.~III ---
the correspondence is $r_d^2 = |F_{tt}|/S_0$ and $2\theta_d = 0\ (\pi)$ for
$F_{tt}>0\ (<0)$ --- with the additional constraints, (ii) and (v) below:

(i) Semileptonic $B$ decays depend on $|V_{ub}/V_{cb}|$ in the same way as in
the Standard Model.

(ii) The ratio $\Delta m_{B}/\Delta m_{B_s}$ depends on $|V_{td}/V_{ts}|$ in
the same way as in the Standard Model.

(iii) The CP asymmetry $a_{\psi K}$ depends on the sign of 
$F_{tt}$~\cite{Buras:2001af}:
\beq\label{apkmfv}
a_{\psi K}={\rm sign}(F_{tt})\sin2\beta \,.
\eeq

(iv) The mass difference $\Delta m_B$ depends on $|F_{tt}|$:
\beq\label{dmbmfv}
\Delta m_B=(\Delta m_B)^{\rm SM}\, {|F_{tt}|\over S_0} \,.
\eeq 
(The QCD correction in MFV models may differ from $\eta_B$, but the
modification is the same for $\Delta m_{B_{d,s}}$ and for the top contribution
to $\epsilon_K$, so it can be absorbed in $F_{tt}$~\cite{Buras:2000xq}.)

(v) The Standard Model contribution to $\epsilon_K$ that is proportional to
Im[$(V_{ts}V_{td}^*)^2$] is multiplied by $F_{tt}$ while the other
contributions remain unchanged:
\beq\label{epsmfv}
\epsilon_K = \epsilon_{tt}^{\rm SM} F_{tt} + \epsilon_{ct}^{\rm SM}
  + \epsilon_{cc}^{\rm SM}.
\eeq

The constraints on $\rhobar$, $\etabar$ and $F_{tt}$ from these processes have
been analyzed in a number of papers (see, for example, 
\cite{Buras:2000dm,Bergmann:2001pm}). In this section we present the MFV
predictions for $A_{\rm SL}$.

The dependence of $A_{\rm SL}$ on new contributions is simple. Since for the
$B^0-\Bbar^0$ mixing amplitude one has $M_{12}=(M_{12})^{\rm SM} F_{tt}/S_0$,
while $\Gamma_{12}$ is not modified, we obtain:
\beq\label{aslmfv}
A_{\rm SL}^{\rm MFV}=A_{\rm SL}^{\rm SM}\, {S_0\over F_{tt}}\,.
\eeq
Thus the size of $A_{\rm SL}$ may be different from the Standard Model.

As concerns the sign of $\ASL$, one might naively think that it could be
opposite to the Standard Model prediction, since ${\rm sign}(A_{\rm SL}^{\rm
MFV}) \propto {\rm sign}(F_{tt})$. However, this is not the case, because 
${\rm sign}(A_{\rm SL}^{\rm SM}) = - {\rm sign}(\etabar)$, and so
\beq\label{signamfv}
{\rm sign}(A_{\rm SL}^{\rm MFV})=-{\rm sign}(\etabar F_{tt}).
\eeq
The product $\etabar F_{tt}$ (in combination with the upper bound on
$|V_{ub}/V_{cb}|$ which implies $\rhobar<1$) determines, however, also the
sign of $a_{\psi K}$:
\beq\label{signmfv}
{\rm sign}(a_{\psi K}^{\rm MFV})={\rm sign}(\etabar F_{tt}),
\eeq
and is therefore experimentally determined to be positive.  We conclude that
there can be no sign difference between the SM and MFV predictions for $A_{\rm
SL}$. In other words, if $A_{\rm SL}$ is experimentally found to be positive,
both the Standard Model and the MFV models will be excluded.

\subsection{Numerical Results}

\begin{figure}
\centerline{\epsfxsize=0.5\textwidth\epsfbox{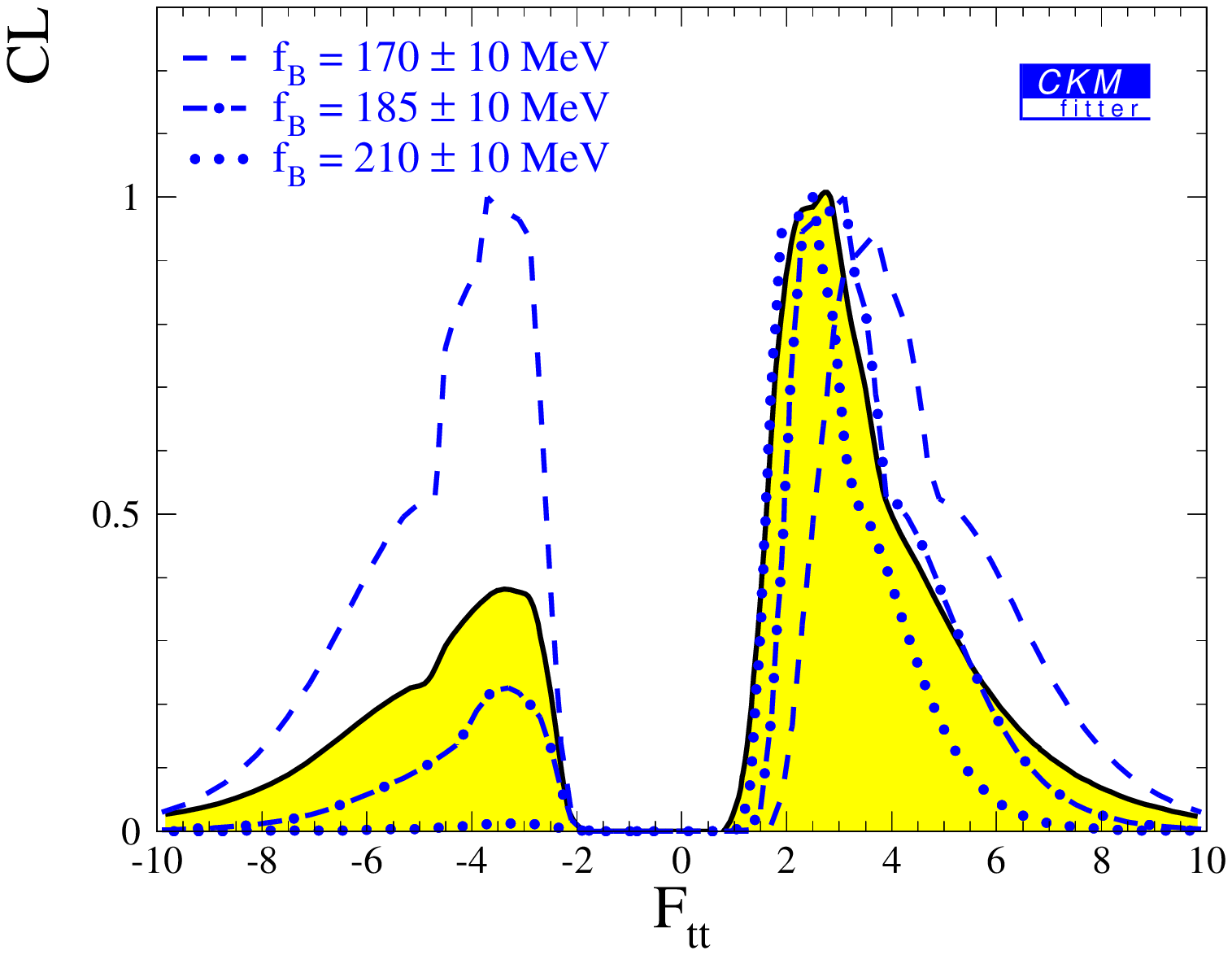} \hfill
  \epsfxsize=0.5\textwidth\epsfbox{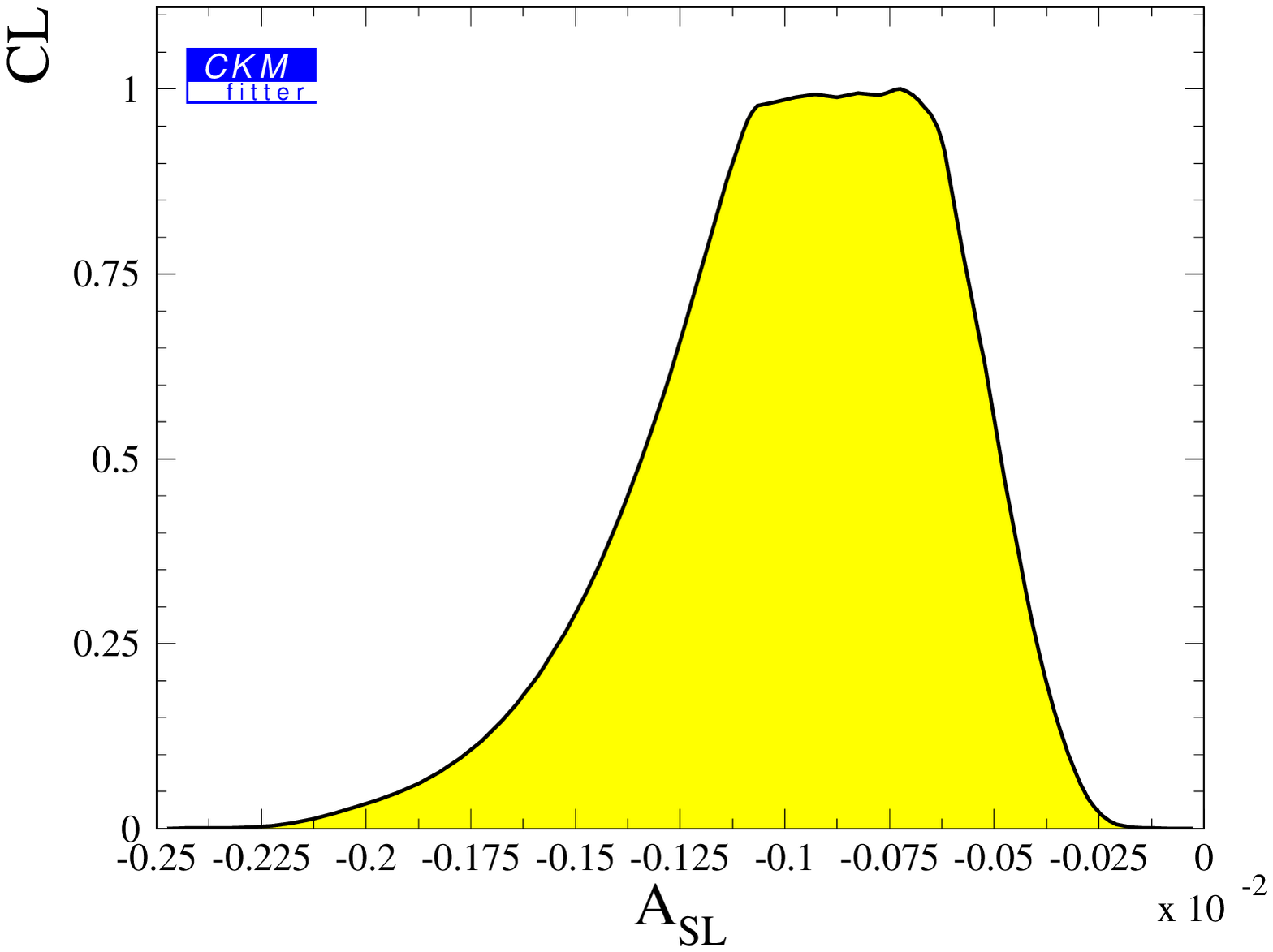} }
\caption{Left: Confidence levels of $F_{tt}$.  The shaded region corresponds to
our ranges for the input parameters (in particular, $f_B=200\pm30\,$MeV).  The
other curves show the sensitivity to $f_B$: $f_B = 170\pm10\,$MeV (dashed),
$f_B = 190\pm10\,$MeV (dash-dotted), and $f_B = 210\pm10\,$MeV (dotted).  In
this last case the $F_{tt}<0$ solution is excluded.  Right: Confidence levels
of $\ASL$ in MFV models.}
\label{fig:ftt1d}
\end{figure}

The shaded region in the left plot in Fig.~\ref{fig:ftt1d} shows the confidence
levels of $F_{tt}$ obtained from the above constraints.  The range of $F_{tt}$
with greater than 10\% CL is
\beq
- 7.3 < F_{tt} < -2.4 \qquad \mbox{and} \qquad 1.2 < F_{tt} < 7.3 \,.
\eeq
The range corresponding to greater than 32\% CL is $-4.3 < F_{tt} < -2.7$ and
$1.4 < F_{tt} < 5.1$.  This implies, using eq.~(\ref{aslmfv}) and
$S_0(m_t^2/m_W^2) \simeq 2.4$, that the magnitude of $\ASL$ can hardly be
enhanced compared to the SM, as it is shown in the right plot in
Fig.~\ref{fig:ftt1d}.  The range of $\ASL$ values with greater than 10\% CL in
MFV models is
\beq
-1.77 \times 10^{-3} < \ASL < -0.33 \times 10^{-3} \,, 
\eeq
while the greater than 32\% CL range is $-1.48 \times 10^{-3} < \ASL < -0.43
\times 10^{-3}$.  Thus, a measurement of $\ASL$ significantly above the SM
prediction would exclude both the Standard Model, and models with minimal
flavor violation.

The existence of the $F_{tt} < 0$ solution is sensitive to the value of $f_B$
(this was first pointed out in Ref.~\cite{Buras:2001af}).  This is shown by the
curves superimposed on the left plot in Fig.~\ref{fig:ftt1d}, corresponding to
$f_B = 170\pm10\,$MeV (dashed), $f_B = 190\pm10\,$MeV (dash-dotted), and $f_B =
210\pm10\,$MeV (dotted).  If future unquenched lattice calculations obtain
$f_B$ above $200\,$MeV with small error, then the confidence level of the
$F_{tt} < 0$ solution is reduced, and this solution practically disappears if
$f_B = 210\pm10\,$MeV or larger.  Note also that if the value of $f_B$
decreases then the CL of the $F_{tt}<0$ solution increases, while that of the
$F_{tt}>0$ solution is reduced.  For example, for $f_B = 170\pm10\,$MeV, both
solutions have about equal CL, and if $f_B = 160\pm10\,$MeV then the CL of the
$F_{tt}>0$ solution is hardly above 50\%.

\begin{figure}
\centerline{\epsfxsize=0.44\textwidth\epsfbox{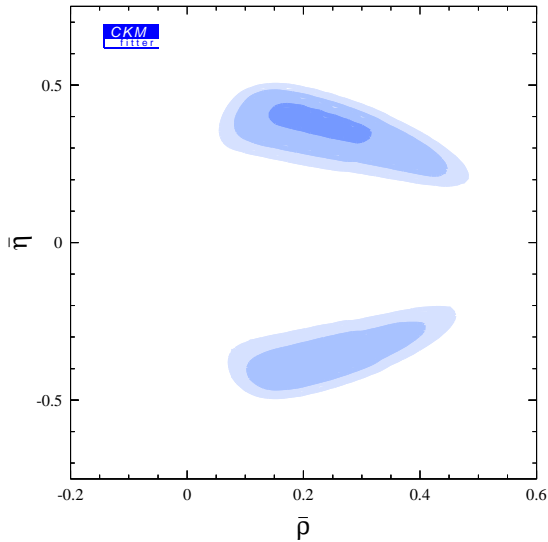}}
\caption{The allowed $\rhobar-\etabar$ range in MFV models.  
The shadings mean the same as in Fig.~\ref{aslsmrhoeta}.}
\label{fig:fttrhoeta}
\end{figure}

Fig.~\ref{fig:fttrhoeta} shows the allowed range of $\rhobar$ and $\etabar$ in
MFV models.  In agreement with Fig.~\ref{fig:ftt1d}, it can be seen that the
$\etabar < 0$ solution is less favored than the allowed $\etabar > 0$ region.

\section{Conclusions}

We investigate $\ASL$, the CP asymmetry in semileptonic $B$ decays, in three
theoretical frameworks.

Within the Standard Model, we improve previous predictions by including
corrections of order $m_c^2/m_b^2$ and $\Lambda_{\rm QCD}/m_b$ [see 
eq.~(\ref{immgacor})]. This leaves the NLO corrections, of order $\alpha_s$, as
the largest source of uncertainty. Our improved calculation gives that the
allowed range of $\ASL$, corresponding to greater than 10\% CL, is:
$-1.3\times10^{-3} < \ASL < -0.5\times10^{-3}$. The smallness of the asymmetry
means that even with improved statistics in the $B$ factories, no useful
constraints on the CKM matrix will arise within the next few years.

Within models of minimal flavor violation, a mild enhancement of the asymmetry 
is possible, $-1.8 \times 10^{-3} < \ASL < -0.3 \times 10^{-3}$. This is again
too small to be observed in the near future. Conversely, if the asymmetry is
measured with a value that is much larger than the SM range, MFV will be
excluded. Moreover, within MFV models, the SM relation, sign($\ASL) =
-$sign($a_{\psi K})$, is maintained; a measurement of a positive $\ASL$ would
therefore exclude both the SM and MFV.

The recent measurements of $\ASL$ do already have meaningful consequences in
probing less constrained extensions of the SM, where there are new sources of
flavor and CP violation. In particular, in models where the new effects can be
neglected in tree-level decays but are significant in flavor changing neutral
current processes, the four observables --- $|V_{ub}|$ from charmless
semileptonic $B$ decay rates, $\Delta m_B$, $a_{\psi K}$ and $\ASL$ ---  depend
on four parameters: the two CKM parameters $\rhobar$ and $\etabar$ and the two
new parameters $r_d^2$ and $2\theta_d$ [see eq.~(\ref{motnp})].  In this
framework, $\ASL$ makes an impact in constraining the allowed range in  the
$r_d^2-2\theta_d$ plane, especially in the region of small $r_d^2$ and large
$\sin2\theta_d$. In this region, there is strong cancellation between the SM
and the new contributions to $B^0-\Bbar^0$ mixing, which  minimizes the
magnitude of the dispersive part and maximizes the relative phase between the
dispersive and absorptive parts. Under these circumstances, $\ASL$ is much
enhanced and opposite in sign to the Standard Model prediction. The recent
experimental results disfavor such a possibility (see 
Fig.~\ref{fig:rdthetadrange}). 

We conclude that improved bounds on $\ASL$ will further constrain new physics 
contributions to $B^0-\Bbar^0$ mixing. If a sizable value of $\ASL$ is measured
in the near future, not only the Standard Model but also its extensions with
minimal flavor violation will be excluded.

\begin{acknowledgments}

We thank  Gerhard Buchalla, Bob Cahn, Fran\c{c}ois Le Diberder, Uli Nierste, 
and especially Andreas H\"ocker for helpful discussions.
Z.L.\ was supported in part by the Director, Office of Science, Office of 
High Energy and Nuclear Physics, Division of High Energy Physics, of the 
U.S.\ Department of Energy under Contract DE-AC03-76SF00098. 
Y.N.\ is supported by the Israel Science Foundation founded by the
Israel Academy of Sciences and Humanities.

\end{acknowledgments}


\end{document}